\documentclass[aps,prl,twocolumn,showpacs]{revtex4}

\usepackage[dvips]{graphicx}
\usepackage{amsmath}
\usepackage{float}
\usepackage{bbm}

\newcommand{\half}{\mbox{$\textstyle \frac{1}{2}$}}
\newcommand{\ket}[1]{\left | \, #1 \right \rangle}

\newcommand{\braket}[2]{\left\langle\, #1\,|\,#2\,\right\rangle}

\newcommand{\av}[1]{\langle #1\rangle}

\newcommand{\vac}{\ket{\textrm{vac}}}

\begin{document}

\title{Generation of twin Fock states via transition from a two-component Mott insulator to a superfluid}

\author{M. Rodr\'iguez, S. R. Clark, and D. Jaksch}
\address{Clarendon Laboratory, University of Oxford, Parks Road, Oxford OX1 3PU, U.K.}

\begin{abstract}
We propose the dynamical creation of twin Fock states, which exhibit
Heisenberg limited interferometric phase sensitivities, in an
optical lattice. In our scheme a two-component Mott insulator with
two bosonic atoms per lattice site is melted into a superfluid. This
process transforms local correlations between hyperfine states of
atom pairs into multi-particle correlations extending over the whole
system. The melting time does not scale with the system size which
makes our scheme experimentally feasible.
\end{abstract} \pacs{03.75.Lm, 03.75.Mn, 03.75.Dg}
\maketitle

\begin{figure}[t]
\includegraphics{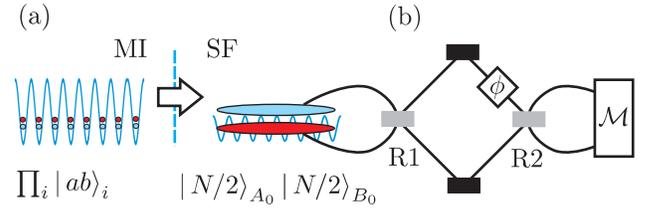}
\caption{(a) Melting of a two-component two-atom MI into a SF. (b)
The internal states $a$ and $b$ represent the arms of an
interferometer with rotations R1, R2 and phase-shift $\phi$ induced
by a Ramsey pulse sequence; $\phi$ is measured by $\mathcal{M}$.}
\label{fig:mzi}
\end{figure}

Degenerate atomic Bose- and Fermi-gases \cite{BEC} provide an
excellent starting point for engineering almost-pure many-particle
quantum states which are an essential resource for novel quantum
technologies. This is illustrated by the realization of a Mott
insulator (MI) \cite{SFMI} in the lowest Bloch band of an optical
lattice which can serve as a quantum memory \cite{spinlattice}.
However, new methods are necessary for attaining some of the most
important $N$-particle states, in particular those which allow a
sensitivity enhancement from the standard quantum limit $\propto
1/\sqrt N$ to the Heisenberg limit $\propto 1/N$ in quantum
metrology \cite{Giovanetti04}. Here we propose a method for
engineering twin Fock states \cite{Holland} starting from a two
component MI with two bosonic atoms per site in an optical lattice
\cite{spinlattice,twoatom}. Our scheme manipulates the hyperfine
states $a$ and $b$ of atom pairs pinned to single lattice sites and
decoupled from one another in the MI regime. In this limit the
dynamics is reduced to a set of identical two particle problems and
correlations between states $a$ and $b$ are accurately induced using
Raman laser pulses or microwaves and collisional interactions
controlled via Feshbach resonances \cite{Feshbach}. These local {\em
two-particle} correlations are then transformed into {\em
multi-particle} correlations extending over the whole system by
melting the MI into a superfluid (SF). This quantum melting can be
experimentally implemented by adiabatically ramping down the depth
of the lattice potential \cite{SFMI} as shown in
Fig.~\ref{fig:mzi}(a). By an appropriate choice of correlations
created in the MI phase a twin Fock state emerges in the resulting
two component superfluid. Using this state in a Mach-Zehnder
interferometer (MZI), shown in Fig.~\ref{fig:mzi}(b), one can
approach sensitivities scaling as $1/N$.

Our method is motivated by theoretical and experimental evidence
\cite{SFMI,Zurek} that long range correlations build up quickly when
melting a MI to a SF. Almost adiabatic melting can be achieved with
ramping times $t_r$ on the order of a few tens of $N/M J_0$ where
$J_0$ is the typical tunneling amplitude between lattice sites
during the melting process and $M$ is the number of lattice sites
\cite{Zurek}. Importantly, in our scheme $M \propto N$ thus
preventing $t_r$ from scaling with $N$. The achievement of
sensitivities at the Heisenberg limit relies acutely on the relative
number of atoms entering each port of the interferometer. This
stringent requirement rules out simple schemes involving either a
$\pi/2$ rotation into a two-component Bose-Einstein condensate (BEC)
or a rapid splitting of a single-component BEC in a double-well
potential, since both result in a binomial relative number
distribution. While number squeezing to a twin Fock state could be
achieved in the latter case by an adiabatic splitting, this
timescale $t_r \propto N$ since $M=2$ is fixed \cite{double,Zurek}.
In contrast, the smaller $t_r$ for our scheme is less demanding on
the suppression of dissipation which is known to degrade the
achievable sensitivities \cite{Barnett,Huelga}. We note another
promising route to generating a twin Fock state through the coherent
dissociation of a molecular BEC as suggested in \cite{Kheruntsyan}.

Our starting point is a MI with two atoms in each lattice site $i$
given by $\ket{\Psi_{ab}}=\prod_i\ket{ab}_i$ \cite{twoatom,MIprep}.
This state, with exactly the same number of $a$ and $b$ atoms, can
be created by collisional interactions involving an auxiliary state
$c$ as experimentally demonstrated in \cite{twoatom}. We analyze the
melting of $\ket{\Psi_{ab}}$, and also of the superposition state
$\ket{\Psi_{aa+bb}} = \prod_i(\ket{aa}_i+\ket{bb}_i)/\sqrt{2}$
obtainable from $\ket{\Psi_{ab}}$ by applying a $\pi/2$-Raman pulse.
For this we use the two component Bose-Hubbard model describing the
dynamics of atoms trapped in the lowest Bloch band of a sufficiently
deep optical lattice. The corresponding Hamiltonian is ($\hbar=1$)
\cite{SFMI}
\begin{eqnarray}
&&\hat{H}=-\sum_{\langle i,j \rangle}(J_a \hat{a}^\dagger_i
\hat{a}_j + J_b\hat{b}^\dagger_i\hat{b}_j) + U \sum_i
\hat{n}_i^a\hat{n}_i^b
\nonumber \\
&&+\frac{V_a}{2} \sum_{i} \hat{n}_i^a(\hat{n}_i^a-1)+\frac{V_b}{2}
\sum_{i} \hat{n}_i^b(\hat{n}_i^b-1),\label{ham}
\end{eqnarray}
where $\hat{a}_{i}(\hat{b}_{i})$ is the bosonic destruction operator
for an $a(b)$-atom localized in lattice site  $i$,
$\hat{n}_i^a=\hat{a}^{\dagger}_i\hat{a}_i$ and
$\hat{n}_i^b=\hat{b}^{\dagger}_i\hat{b}_i$, while $\langle i,j
\rangle$ denotes summation over nearest-neighbors. The parameter
$J_{a(b)}$ is the tunneling matrix element for atoms in state
$a$($b$); $V_{a(b)}$ and $U$ are the on-site intra- and
inter-species interaction matrix elements respectively. For
simplicity we only consider the symmetric case with $J_{a(b)}=J$,
$V_{a(b)}=V$. The ratio between the matrix elements is determined by
the lattice depth \cite{SFMI} and additionally $U$, $V$ can be
controlled via Feshbach resonances \cite{Feshbach} or by shifting
the $a$ and $b$ atoms away from each other using state-dependent
lattices \cite{spinlattice}. We denote the total number of atoms in
state $a(b)$ as $\hat{N}_{a(b)}= \sum_i \hat{n}^{a(b)}_i $ and
introduce Schwinger boson operators
$\hat{J}_x=\sum_i(\hat{a}_i^\dagger\hat{b}_i+\hat{b}_i^\dagger\hat{a}_i)/2$,
$\hat{J}_y=\sum_i(\hat{a}_i^\dagger\hat{b}_i-\hat{a}_i^\dagger\hat{b}_i)/2\textrm{i}$,
$\hat{J}_z=(\hat{N}^a - \hat{N}^b)/2$ and ${\bf \hat{J}} =
(\hat{J_x},\hat{J_y},\hat{J_z})$.

We illustrate the outcome of melting two-component MI ground
states with fixed total particle number $N=2M$ assuming adiabatic
evolution in an $M$-site system with periodic boundary conditions.
The melting starts deep in the MI regime, where $J$ can be
neglected. For $U < V$ the nondegenerate MI ground state is
$\ket{\Psi_{ab}}$ and this adiabatically melts into the
nondegenerate SF ground state
$\ket{\Psi_{\textrm{tf}}}=\ket{N/2}_{A_0}\ket{N/2}_{B_0} \propto
(\hat{A_0}^{\dagger})^{N/2} (\hat{B_0}^{\dagger})^{N/2} \vac$ for
$V/J \rightarrow 0$. Here all $N/2$ $a$-atoms are in the same
delocalized symmetric mode $\hat{A_0} \propto \sum_i \hat{a}_i$,
all $N/2$ $b$-atoms are in mode $\hat{B_0}\propto \sum_i
\hat{b}_i$ and $\vac$ is the vacuum state. Thus, adiabatic melting
of $\ket{\Psi_{ab}}$ provides a direct means of obtaining a
twin Fock state $\ket{\Psi_{\textrm{tf}}}$ with zero relative atom
number difference.

In the opposite case $U > V$ the system exhibits spatial separation
of the $a$ and $b$ components and has a large number of degenerate
ground states. For an even number $N_a$ of $a$-atoms this degeneracy
is lifted by completely connected hopping \cite{Fisher} and the
state $\ket{\Psi^{N_a}_{\textrm{sep}}}=\mathbbm{S}\{\prod_{(i \leq
N_a/2)}\ket{aa}_i\prod_{(N_a/2<j\leq N)}\ket{bb}_{j}\}$ is
adiabatically connected to the nondegenerate SF ground state
$\ket{N_a}_{A_0}\ket{N-N_a}_{B_0}$, where $\mathbbm{S}$ denotes
symmetrization over lattice site configurations. Our second initial
state $\ket{\Psi_{aa+bb}}$ is a binomial superposition of ground
states $\ket{\Psi^{N_a}_{\textrm{sep}}}$ and therefore results in
the melted state
\begin{equation}
\ket{\Psi_{\textrm{mac}}}=\frac{1}{\sqrt{2^{N/2}}}\sum_{m=0}^{N/2}\left(
\begin{array}{c} N/2 \\ m \end{array} \right)^{1/2}
\ket{2m}_{A_0}\ket{N-2m}_{B_0}.\nonumber
\end{equation}
After rotation, the overlap of this state with the macroscopic
superposition state \cite{Heinzen} $\ket{\Psi_{\textrm{max}}} =
(\ket{N}_{A_0}\ket{0}_{B_0} + \ket{0}_{A_0}\ket{N}_{B_0})/\sqrt{2}$
is found to be ${\cal O}=|\langle\Psi_{\textrm{max}}|\exp(\textrm{i}
{\pi} \hat{J}_y /2)|\Psi_{\textrm{mac}}\rangle| = \sqrt[4]{8/9} >
0.97$ for $N \rightarrow \infty$ and this limit is monotonically
attained with ${\cal O}>0.97$ for $N>20$.

\begin{figure}[t]
\includegraphics{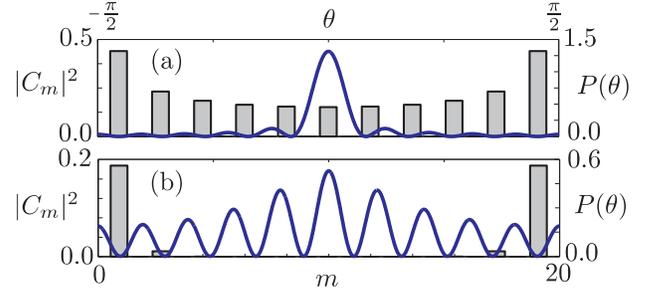}
\caption{The amplitudes $|C_m|^2$ (left-bottom axes) and
corresponding relative phase probability distribution $P(\theta)$
(right-top axes) after the rotation R1 of (a)
$\ket{\Psi_{\textrm{tf}}}$ and (b) $\ket{\Psi_{\textrm{mac}}}$ with
$N=20$.} \label{fig:states}
\end{figure}

We can use either of the final states obtained by adiabatic
melting to realize sensitivities $\propto 1/N$ in the MZI setup
shown in Fig.~\ref{fig:mzi}(b). For the twin Fock state
$\ket{\Psi_{\textrm{tf}}}$ R1 implements the conventional
beam-splitter operation $\exp(\textrm{i} {\pi} \hat{J}_x /2)$. In
the case of $\ket{\Psi_{\textrm{mac}}}$ R1 rotates the state
according to $\exp(\textrm{i} {\pi} \hat{J}_y /2)$ as discussed
above. Then a relative phase $\phi$ is induced in one of the arms
of the MZI. The operation R1 can be implemented by rapid resonant
$\pi/2$-Raman pulses, while the relative phase $\phi$ could be
induced by an appropriate off-resonant pulse. The achievable phase
sensitivity inside the interferometer (after R1) can be computed
from the relative phase probability distribution $P(\theta)$. For
a given state $\ket{\psi} = \sum_{m=0}^{N} C_m
\ket{m}_{A_0}\ket{N-m}_{B_0}$ this is computed as $P_s(\Delta
\theta) = |\sum_{m=0}^{N} C_m \exp(-im\Delta\theta)|^2/(s+1)$
where $\Delta \theta$ is a multiple of $2\pi/(s+1)$. The
distribution $P( \theta)$ is obtained by multiplying $P_s(\Delta
\theta)$ with $(s+1)/2\pi$ and taking the limit $s \rightarrow
\infty$ \cite{Barnett}. As shown in Fig.~\ref{fig:states}(a)
$P(\theta)$ is sharply peaked for the rotated
$\ket{\Psi_{\textrm{tf}}}$ and Fig.~\ref{fig:states}(b) displays
the oscillatory behavior of $P(\theta)$ for the rotated
$\ket{\Psi_{\textrm{mac}}}$. The sensitivity $\delta \phi$ of each
state is quantified by the half width at half maximum (HWHM) of
$P(\theta)$ around some fixed phase typically taken as $\theta =
0$. From this definition it can be shown for both states that
$\delta \phi \propto 1/N$ and so scale at the Heisenberg limit
\cite{Barnett,Holland,Heinzen}. The final step in the MZI consists
of rotation $R_2$ and measurement $\cal M$ of a phase dependent
quantity which exhibits this Heisenberg limited phase resolution.
For instance the measurement of parity as discussed in
\cite{Campos03} or $\hat{J}^2_z$ as proposed in
\cite{Dunningham04} can achieve this. These methods exploit
nonlinearities caused by atomic collisions and are thus realizable
in our setup.

\begin{figure}[t]
\includegraphics{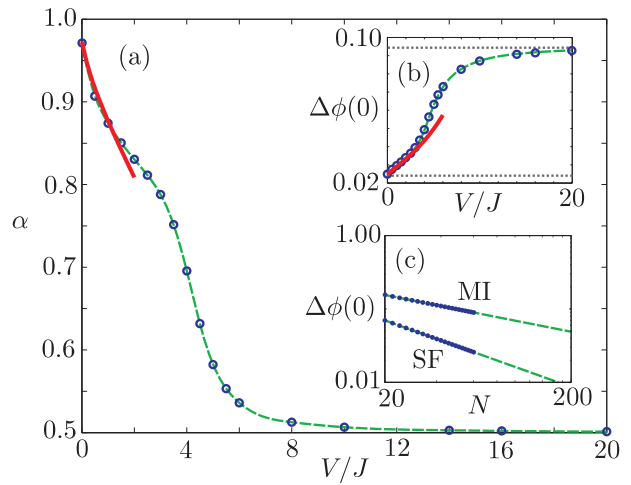}
\caption{(a) Scaling $\alpha$ in $\Delta \phi \sim N^{-\alpha}$
against $V/J$ of ground states of $\hat H$ with $U=0$. The scaling
was obtained from calculations with $N=20$ up to $N=60$ with $M=N/2$
for both the 1D numerical results $(\circ)$ in a box system and the
$N$-conserving Bogoliubov results (solid line). (b) Sensitivity
$\Delta \phi$ as a function of $V/J$ for $N=60$. The dotted lines
show exact results for the twin Fock SF (bottom) and the MI (top)
states, while the thick solid line shows the Bogoliubov result. (c)
$\Delta \phi$ as a function of $N$ for $V/J=20$ (MI) and $V/J=1/2$
(SF) giving $\alpha$ of the corresponding points in (a). In all
cases the dashed lines are to guide the eye.} \label{fig:scaling}
\end{figure}

We now analyze the achievable sensitivity for incomplete melting at
finite (residual) values of $V/J$ before considering experimental
imperfections arising from non-adiabatic ramping and particle loss.
 We restrict our considerations to the measurement
of $\hat{J}^2_z$ and to the twin Fock state due to its greater
experimental feasibility. We study these effects in terms of the
noise $\Delta \phi$ on the $\phi$ dependent observable
$\hat{J}^2_z$. Error propagation theory gives $\Delta \phi={\Delta
\hat{J}^2_z}/{|\partial \langle \hat{J}^2_z\rangle/\partial \phi|}$
where $\langle \hat{J}^2_z\rangle$ and $\Delta \hat{J}^2_z$ are the
average and the spread respectively. For states which are zero
eigenvectors of $\hat{J}_z$, such as $\ket{\Psi_{\textrm{tf}}}$,
this gives
\begin{equation}
\Delta \phi^2=\frac{\sin^2\phi(\langle \hat{J}_x^4 \rangle-\langle
\hat{J}_x^2 \rangle^2)+ \cos^2 \phi \langle \hat{J}_x \hat{J}_z^2
\hat{J}_x \rangle}{4 \cos^2 \phi \langle \hat{J}_x^2 \rangle^2}.
\end{equation}
At $\phi=0$ the sensitivity reduces to $\Delta
\phi({0})=\half\langle \hat{J}_x^2 \rangle^{-1/2}$ which can be
expressed entirely in terms of the one-particle density matrices
${\rho}^{a}_{ij} = \langle \hat{a}^{\dagger}_i\hat{a}_j \rangle$ and
${\rho}^{b}_{ij} = \langle \hat{b}^{\dagger}_i\hat{b}_j \rangle$
using $\langle \hat{J}_x^2 \rangle=[\sum_i(\rho^a_{ii} +
\rho^b_{ii}) + \sum_{i,j}(\rho^a_{ij}\rho^b_{ji} +
\textrm{h.c.})]/4$. The initial MI state $\ket{\Psi_{ab}}$ with no
off-diagonal correlations ${\rho}^a_{ij}={\rho}^b_{ij}=\delta_{ij}$
yields the standard quantum limit $\Delta \phi=1/{\sqrt{2 N}}$
(cf.~Fig.~\ref{fig:scaling}(a) at $V/J\gg 1$). The final SF
$\ket{\Psi_{\textrm{sf}}}$ with long-range correlations
$\rho^a_{ij}=\rho^b_{ij}=1$ asymptotically recovers the Heisenberg
limit $\Delta \phi=1/{\sqrt{N^2/2 + N}}$
(cf.~Fig.~\ref{fig:scaling}(a) at $V/J=0$). Thus the scaling $\Delta
\phi \propto N^{-\alpha}$ changes from $\alpha=1/2$ to $\alpha=1$
during the melting.

The presence of a residual intra-species interaction $V$ results in
quantum depletion of the populations in the $A_0$ and $B_0$ modes
which reduces the attainable sensitivity. We consider this effect
for $U=0$, achieved for example by fully separating the sites of a
spin-dependent lattice. In the SF regime $J \gg V$ a translationally
invariant system is well described by the $N$-conserving Bogoliubov
wavefunction
\begin{equation}
\ket{\psi_{\textrm{bog}}}\sim (\hat{\Lambda}_a)^{N/2}
(\hat{\Lambda}_b)^{N/2}\ket{\rm vac}, \label{bog}
\end{equation}
where $\hat{\Lambda}_a = (\hat{A}_0^{\dagger 2}-\mathop \sum_{{\bf
q} \neq 0} c_{\bf q} \hat{A}_{\bf q}^\dagger \hat{A}_{-{\bf
q}}^\dagger)$ with $\hat{A}_{\bf q}$ the ${\bf q}$ quasi-momentum
modes for $a$, and similarly $\hat{\Lambda}_b$ is defined in terms
of $\hat{B}_{\bf q}$ for component $b$. Within this ansatz the
depletion is identical for both components and is given by
$\zeta=\sum_{{\bf q}\neq0} n_{\bf q} = c_{\bf q}^2/(1-c_{\bf q}^2)$
where $n_{\bf q}=\av{\hat{A}^{\dagger}_{\bf q}\hat{A}_{\bf
q}}=\av{\hat{B}^{\dagger}_{\bf q}\hat{B}_{\bf q}}$ and the
amplitudes $c_{\bf q} \ll 1$ can be solved in terms of $V/J$ and are
given in \cite{Bogoliubov}. The sensitivity is then $\Delta
\phi({0})={1}/{{[N^2(1-\zeta)^2/2 +N +2\sum_{{\bf q}\neq 0} n_{\bf
q}^2}]}^{1/2}$. For a three dimensional (3D) optical lattice the
depletion $\zeta=(V/J)^{3/2} / 3 \pi^2 \sqrt{2} $ and the sum
$\sum_{{\bf q}\neq 0} n_{\bf q}^2=(V/J)^{3/2}(3\pi-8)/12 \sqrt{2}$
are constant in $N$. Within the region of validity of the Bogoliubov
ansatz there will thus be no degradation of the sensitivity scaling
$\alpha$. Similarly we find no decrease of $\alpha$ in 2D within the
Bogoliubov ansatz. In contrast, for a 1D system the depletion
increases with $N$ since there is no true condensate for non-zero
$V$ and long-range correlations decay algebraically. Consequently
the scaling $\alpha$ decreases with increasing $V/J$ as shown in
Fig.~\ref{fig:scaling}(a) for a finite number of particles.

We explore the sensitivity scaling between the SF and MI limit in 1D
numerically using the Time-Evolving-Block-Decimation algorithm
\cite{TEBD}. For specific values of $N$ we compute the ground states
of the Hamiltonian Eq.~(\ref{ham}) over a range of $V/J$ and $U=0$
and calculate their sensitivities. The result for $N=60$ in
Fig.~\ref{fig:scaling}(b) shows a smooth transition between the
ideal SF and MI results and agrees with the Bogoliubov result for
$V/J \leq 2$. By repeating this calculation for different $N$ we
extract the sensitivity scaling $\alpha$. For all values of $V/J$
the numerical results are well approximated by a power-law $\Delta
\phi \sim N^{-\alpha}$ as shown for the extreme cases in
Fig.~\ref{fig:scaling}(c). We combine these results in
Fig.~\ref{fig:scaling}(a) which displays the scaling $\alpha$ as a
function of $V/J$. The value of $\alpha$ decreases smoothly from the
Heisenberg limit in the SF regime to the standard quantum limit in
the MI regime. As also shown in Fig.~\ref{fig:scaling}(a) the
1D-Bogoliubov ansatz Eq.~(\ref{bog}) is consistent with the numerics
for $V/J \leq 2$ and the values of $N$ used here. In the MI regime a
number conserving particle-hole ansatz \cite{Zurekph} underestimates
long-range correlations and therefore does not predict an increase
of $\alpha$ from $1/2$ at finite $V/J$. This emphasizes the
importance of the growth of long range correlations for increasing
$\alpha$ above the standard quantum limit. While our numerical
calculations are performed for sizes of 1D optical lattices
currently used in experiments \cite{SFMI} we expect the scaling
$\alpha$ to drop from $\alpha=1$ to $\alpha=1/2$ at vanishingly
small $V/J$ when $N \rightarrow \infty$ because of the lack of true
long range correlations. In 2D and 3D this reduction is instead
expected to occur at finite $V/J$ close to the phase transition
point $(V/J)_{\rm crit}$ below which long range correlations exist
at large $N$.

\begin{figure}[t]
\includegraphics{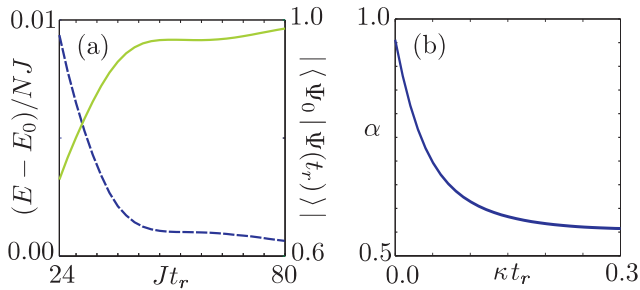}
\caption{(a) Energy difference per atom $(E-E_0)/NJ$ between
$\ket{\Psi(t_r)}$ and the final SF ground state $\ket{\Psi_0}$ (left
axis, dashed lines) and overlap $|\braket{\Psi(t_r)}{\Psi_0}|$
(right axis, solid line) as a function of the ramping time $t_r$.
Starting from the MI ground state at $V_0/J=20$ we ramped $V(t)$
down to a final $V_{t_r}/J=1/2$ using different ramping times $t_r$
and $U=0$, $N=50$ and $M=25$. (b) Scaling $\alpha$ of $\delta \phi$
with $N$ for $\ket{\Psi_{\textrm{tf}}}$ under the presence of
atom-loss at a rate $\kappa$ obtained from results for $N=20$ up to
$N=60$. Note that $\alpha$ does not reach unity because of the
finite $N$ used.} \label{fig:adiabatic}
\end{figure}

In experiments the ramping time $t_r$ will be finite and so the
melting is never perfectly adiabatic. We numerically compute the
dynamical ramping of $V/J$ for the Hamiltonian Eq.~(\ref{ham}) with
$U=0$ from the MI ground state. In Fig.~\ref{fig:adiabatic}(a) the
energy difference per atom between the final ramped state
$\ket{\Psi(t_r)}$ and the SF ground state $\ket{\Psi_0}$ is shown as
a function of the ramping time $t_r$, along with the corresponding
many-body overlap $|\braket{\Psi(t_r)}{\Psi_0}|$. For the infinite
Bose-Hubbard-model it has been shown \cite{Zurek} that $t_r \gg
V_{\textrm{max}}/J^2$ guarantees adiabatic evolution. Our numerics
of a finite-sized system agree with this result giving a near-unit
overlap and small energy differences for ramping times of $t_r \sim
3 V_0/J^2$ with $V_0$ the initial interaction strength in the MI.

Given that the ramping time $t_r$ is much longer than the time
required to implement the MZI in Fig.~\ref{fig:mzi}(b) we consider
the influence of atom loss over a time $t_r$. The build-up of long
range correlations during the melting process has been found to be
robust to atom loss processes \cite{SFMI}. However, atom loss will
cause uncertainty in the relative number of $a$ and $b$ atoms in the
final state $\ket{\Psi_{\textrm{tf}}}$. We use the model introduced
in \cite{Barnett} to calculate the sensitivity scaling $\alpha$
assuming the state $\ket{\Psi_{\textrm{tf}}}$ is subject to atom
loss at a rate $\kappa$ for time $t_r$. As shown in
Fig.~\ref{fig:adiabatic}(b) we find that $\alpha \approx 1$ for
$\kappa t_r \ll 0.1$. In combination with the adiabaticity condition
on $t_r$ Heisenberg limited sensitivities can thus be achieved for
$\kappa \ll J^2/30 V_0$. This condition is within the reach of
current optical lattice technology where tunneling amplitudes $J$ of
a few hundred Hz and loss rates on the order of Hz can be achieved
\cite{SFMI}. Furthermore, additional fluctuations in the overall
particle number $N$ which occur from shot to shot do not affect the
scaling $\alpha$ \cite{Dunningham04}. In the experiment a harmonic
trapping potential is likely to be present and causes an outer shell
of singly occupied lattice sites with atoms in the auxiliary state
$c$. Additionally, an imperfect Raman transition during the creation
of the initial MI state may leave a $c$ atom pair within the
$\ket{\Psi_{ab}}$ state. In both cases the presence of a small
number of $c$-atom impurities will not destroy the coherence of the
final SF state, as seen in recent experiments \cite{Gunter}. Also
these atoms will remain in state $c$ throughout the whole process
and thus not contribute to the particle number fluctuations
\cite{twoatom}.

In summary we have shown that quantum melting of a two-component MI
in an optical lattice provides a viable route to engineer twin Fock
on timescales $t_r$ smaller than current experimental atom-loss
times. In particular the creation time $t_r$ does not scale with
$N$. Our scheme also exploits the accurate controllability of
two-atoms in a single lattice sites to minimize fluctuations in the
relative numbers of $a$ and $b$ atoms which is crucial for achieving
Heisenberg limited sensitivities $\propto 1/N$.

We acknowledge fruitful discussions with I.~Bloch and C.~Simon. This
work is supported by the EPSRC project EP/C51933/1 and the EU
project OLAQUI.

\end{document}